# Fluorescence lifetime imaging via spatio-temporal speckle patterns in single-pixel camera configuration

**J. JUNEK,**[1,2,*] **K. ŽÍDEK**[1]

[1]*Regional Center for Special Optics and Optoelectronic Systems (TOPTEC), Institute of Plasma Physics, Czech Academy of Science v.v.i., Za Slovankou 1782/3, 182 00 Prague 8, Czech Republic*
[2]*Technical University in Liberec, Faculty of Mechatronics, Informatics and Interdisciplinary Studies, Studentská 1402/2, 461 17 Liberec, Czech Republic*
[*] *Author to whom correspondence should be addressed. Email: junekj@ipp.cas.cz*

**Abstract:** Photoluminesce (PL) spectroscopy offers excellent methods for mapping the PL decay on the nanosecond time scale. However, capturing maps of emission dynamics on the microsecond time scale can be highly time-consuming. We present a new approach to fluorescence lifetime imaging FLIM, which combines the concept of random temporal speckles excitation (RATS) with the concept of a single-pixel camera based on spatial speckles. The spatio-temporal speckle pattern makes it possible to map PL dynamics with unmatched simplicity. Moreover, the method can acquire all data necessary to map PL decays in the microsecond timescale within minutes. We present proof of principle measurements for two samples and compare the reconstructed decays to the non-imaging measurements. Finally, we discuss the effect of the preprocessing routine and other factors on the reconstruction noise level. The presented method is suitable for lifetime imaging processes in several samples, including charge carrier dynamics in perovskites or monitoring solid-state luminophores with a long lifetime of PL.

## 1. Introduction

Fluorescence lifetime imaging (FLIM) is an essential spectroscopic method in various fields, including medicine, biology, and material science. Interest in FLIM and its broad applicability stimulate its development. Therefore, FLIM has many different implementations based on a variety of fundamental methods for measuring photoluminescence (PL) decay. One can find gated photoluminescence counting [1], streak camera [2], time-domain analog recording technique [3], or frequency-domain analog recording technique [4]. However, the most commonly used method is time-correlated single-photon counting (TCSPC) [5,6].
The TCSPC is a powerful method to trace PL decay with the lifetime in the nanosecond time scale. However, due to the principle of TCSPC operation, data acquisition can take several hours for samples with a PL lifetime in the microsecond timescale. Therefore, the reduction of the acquisition time in FLIM has become a topic on its right discussed in the literature [7]. A possible way to reduce the acquisition time is to apply to the so-called compressed sensing, where the image can be reconstructed from a highly reduced dataset [8, 9]. However, these works rely on the TCSPC and, despite reducing the acquisition times, FLIM of the samples with the long-lived PL decay still represents an issue. It is also worth noting that the standard FLIM methods are typically costly setups.
In this paper, we present an entirely new concept of FLIM. The concept is based on the use of speckle patterns, both in the spatial and temporal sense, to map the PL decay of a sample. We combine compressive imaging, namely the concept of a speckle-based single-pixel camera [10], with our recent work [11], where random temporal speckles (abbr. RATS) make it possible to trace PL dynamics. In the presented concept, we employ spatio-temporal speckles, which are generated by using two diffusers. The speckles can be generated with any coherent excitation source, i.e. without the need for a pulsed laser. At the same time, the detector is a standard single-pixel detector, e.g. a photomultiplier. The setup is, therefore, very simple, robust and low-cost. Owing to the novel approach to PL decay acquisition, the method is highly suitable for mapping the PL dynamics on the microsecond timescale, where the FLIM dataset can be acquired on the scales of minutes. We demonstrate this on proof-of-principle measurements by imaging PL decay of selected scenes (color filters and Si nanocrystal layers) and we also discuss the effect of speckle properties on the resulting noise level.

The presented method can serve as a simple approach to characterize the morphology of samples with prominent PL decay on the microsecond timescale, which includes halide perovskite samples, solid-state luminophores or Si nanocrystals [12-14].

## 2. Principle of the method

The cornerstone of the concept is the RATS method, which is described in detail in our previous work [11]. This novel method for the measurement of PL decay uses randomly fluctuating intensity $I_{EXC}$ to excite a sample. PL signal $I_{PL}$ is then given as a convolution of $I_{EXC}$ and PL decay $I_D$, according to Eq. (1).

$$I_{PL} = I_{EXC} * I_D \qquad (1)$$

Therefore, $I_D$ can be extracted via the convolution theorem using Fourier transform. The random character of $I_{EXC}$ allows us to measure in a broad range of frequencies. Thus, a single dataset is enough information for proper $I_D$ reconstruction. Since the original method acquires PL decay only for a single spot, we will hereafter denote the method as 0D-RATS.

An efficient approach to converting the 0D-RATS method to the imaging mode is to use the single-pixel camera. The principle of a single-pixel camera can be seen in many review articles [15]. In an SPC experiment, the measured sample is illuminated by a set of masks see (Fig. 1), which were speckle patterns in our experiment. Each illuminating random mask excites PL in different parts of the sample. After illuminating the sample with a sufficient number of masks, it is possible to retrieve the spatial information by detecting the overall level of the emitted PL and by using dedicated algorithms, as we will describe in the following paragraphs. However, the condition that must always be met is the linear dependence between the measured and the reconstructed data.

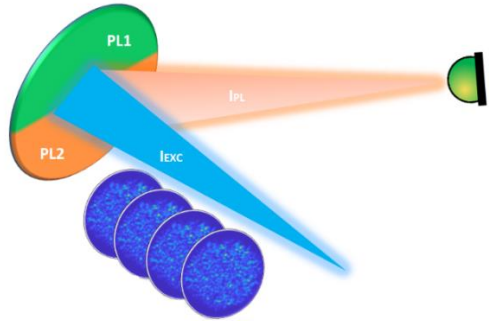

Fig.1. Scheme of the single-pixel camera image acquisition by using speckle patterns – see text for details.

In our FLIM approach, the illuminating masks are generated with a movable diffuser, which is placed behind the source of light with randomly fluctuating intensity in time, i.e. temporal speckles. Thus, we attain spatio-temporal speckles, which retain the same spatial pattern $S(x,y)$, while the overall intensity is rapidly blinking as $I_{EXC}(t)$. In other words, the measured sample is illuminated with a blinking pattern $P(x,y,t) = S(x,y).I_{EXC}(t)$. To map the PL decay $I_D$, i.e.

retrieve $I_D(x,y,t)$, it is necessary to detect $I_{EXC}(t)$ with a diode, $I_{PL}(t)$ with a photomultiplier, and the speckle pattern $S(x,y)$ with an array 2D detector (e.g. CMOS camera).

The Eq. (1) can be rewritten for a 2D sample into a more general case, where $n$ areas with different $I_D$ are measured. The total emitted $I_{PL}$ is the sum of the contributions from all sample spots:

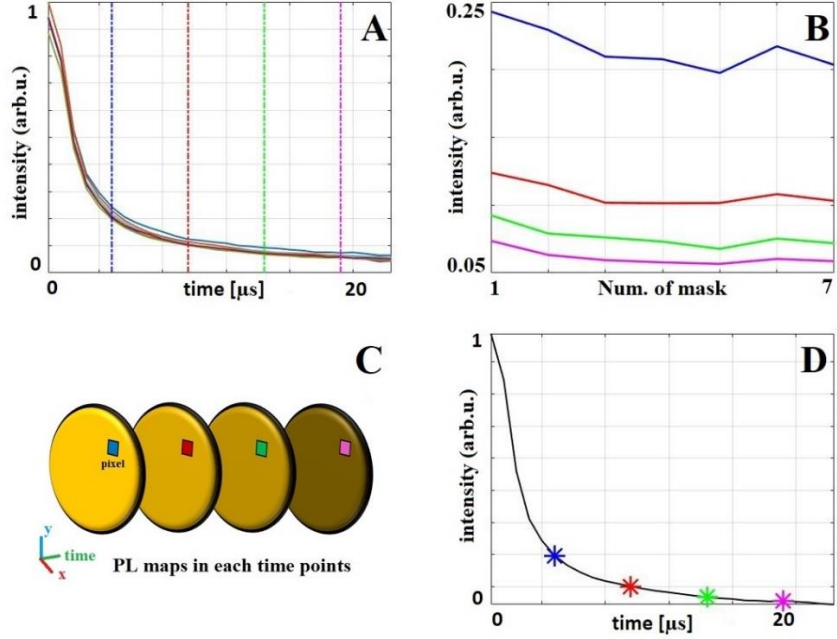

Fig. 2. (A) Examples of simulated $I_D$ curves, which differ due to the change of the illuminating mask. (B) Varying intensity for the seven different masks in selected time-points of simulated $I_D$ curves – see dashed lines in panel A. (C) Example of reconstructed PL maps. Selected pixels are marked with a color corresponding to the time point. (D) A simulated example of reconstructed $I_D$ in four time points. Reconstruction at multiple points would copy the entire $I_D$ (solid black line).

$$\sum_{i=1}^{n} I_{PL(i)} = \sum_{i=1}^{n} \mathbb{F}(I_{EXC}) \mathbb{F}(I_{D(i)}) \qquad (2)$$

For a given mask, we can evaluate from measured data an average PL decay of the entire illuminated area $I_{DA}$:

$$I_{DA} = \mathrm{Re}\left\{ \mathbb{F}^{-1}\left[ \frac{\mathbb{F}\left(\sum_{i=1}^{n} I_{PL(i)}\right)}{\mathbb{F}(I_{EXC})} \right] \right\} = \mathrm{Re}\left\{ \mathbb{F}^{-1}\left[ \frac{\left(\sum_{i=1}^{n} \mathbb{F}(I_{EXC}) \mathbb{F}(I_{D(i)})\right)}{\mathbb{F}(I_{EXC})} \right] \right\} \qquad (3)$$

A different $I_{DA}$ will be detected for each mask, as it is illustrated in Fig. 2A. The variation of the $I_{DA}$ value, i.e. PL decay, for a selected time and different masks will be denoted as $d$. See Fig. 2B for an example of a short subset of seven masks and four different times. We can vectorize each illumination mask into a single row of the so-called observing matrix $A$ and we can also vectorize the map of the PL intensity into a vector $m$. In this case, we can express their relation as a simple matrix multiplication:

$$d = Am \qquad (4)$$

Matrix *A* has the number of the column corresponding to the number of map pixels *N*, while the number of rows follows the number of used masks *M*, i.e. number of measurement. The ratio between *N* and *M* determines a compressed ratio $k = N/M$.

We aim at solving an underdetermined system, which can be accomplished by using a compressed sensing algorithm, where we employ a regularization. In this work, we used algorithm TVAL3 based on the minimization of total variation *TV* of reconstructed images and following Eq. (5) [16].

$$\min \left\{ \|d - Am\|_2^2 + TV(m) \right\} \qquad (5)$$

By using Eq. (5), it is possible to reconstruct the PL map $m(x, y)$ for each time point t (see Fig. 2C). By using the knowledge, that we reconstruct a PL image, we can constrain the solution as $m \in R$ and m≥0. If we stack the individual $m(x, y)$ behind each other, we create a 3D matrix $m(x, y, t)$ which corresponds to the appropriate PL intensity $I_D$ (x,y,t) of the sample. Therefore, we can also extract PL decay for any selected spot of the sample (see Fig. 2D) and we can fit the obtained PL dynamics $I_D$ with a single or multi-exponential decay to get the lifetime map.

## 3. Optical setup

The used optical setup is depicted in Fig. 3. We used a CW laser at wavelength 405 nm (IO matchbox laser diode, Free-space) as a light source. The combination of focusing lens *A* (f = 25,4 mm), rotating diffuser (average grain size 3,87 um) with collimating lens *B* (f = 75 mm), and an aperture (diameter 1.5 mm) generated intensity randomly fluctuating in time $I_{EXC}(t)$.

The generation of a random mask (spatial speckles) is achieved with another focusing lens *C* (f = 25.4 mm). The beam is focused on a movable diffuser (average grain size 8,06 um) and the diffused light is again collimated with lens D (f = 50 mm). The resulting mask pattern was blinking, according to $I_{EXC}(t)$.

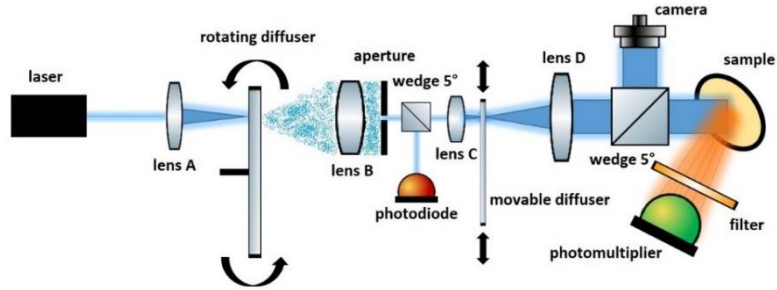

Fig. 3. Scheme of the used optical setup – see text for details.

The patterned beam is split twice with two N-BK7 glass wedges (5 °), which reflect about 6% of the incident intensity. The first reflection is used to detect $I_{EXC}(t)$ with a Si amplified photodetector (Thorlabs PDA8A2, rise time 7 ns). The second reflected beam is used to acquires the mask pattern with a camera (CMOS, IDS UI-3240ML-M-GL). The transmitted pattern is used to illuminate the measured sample. PL emitted from the excited sample, i.e. $I_{PL}(t)$ signal, was detected with a Hamamatsu photomultiplier (PMT) module type H10721-20 (rise time 0.6 ns). The scattered excitation light was blocked by a cut-off filter at 500 nm (Thorlabs, FEL0500). The detected PL signal was amplified by an SRS amplifier model SR445A and read out by a TiePie Handyscope HS5-110XM USB oscilloscope.

The laser beam intensity entering the setup is 138.5 mW, while the full average intensity that illuminates the measured sample oscillates around 5.5 µW. The overall efficiency of the system is about 0.003 %. The size of the measured area size was about 18 mm² and is given by the size of the generated speckle masks. It is possible to scale the field of view by adjusting the collimating lens D.

## 4. Results and discussion

As a proof-of-principle experiment, we carried out imaging of a combination of orange absorbing cut off filter OG565 and Si wafer with a nanoporous surface prepared by electrochemical etching of Si wafer in hydrofluoric acid and ethanol solution [14, 17]. Both samples were measured previously by the 0D-RATS method and the resulting PL decay shapes were verified with a standard method, namely streak camera [11].

### 4.1 Single-pixel camera PL map reconstruction

Our method is based on compressive imaging and requires iterative image reconstruction, which was described in Eq. (5). The crucial parameters (together with their set value) were: *mu* ($2^9$), *beta* ($2^6$). The reconstruction parameters were set according to the reconstruction of the testing experiments and simulations and the same parameters are used for all the presented images.

### 4.2. Mask preprocessing

We captured the speckle mask on a CMOS chip and, prior to its use, we carried out a set of operations to convert the speckle image into a suitable form for our calculation. The first part of preprocessing was the cropping of the mask. Since the mask did not occupy the entire camera chip, the image was cropped so that the information value remains and at the same time we reduce the number of reconstructed pixels $N$.

The second part was the mask rescaling. A laser speckle pattern is a natural random pattern, where the dimensions of each speckle vary around a certain mean value. For this reason, it is unclear how the high-resolution camera image of laser speckles $a_M$ should be rescaled into the image $a_M'$ used in the measurement matrix $A$ while retaining the useful information. An example of such rescaling is present in Fig. 4.

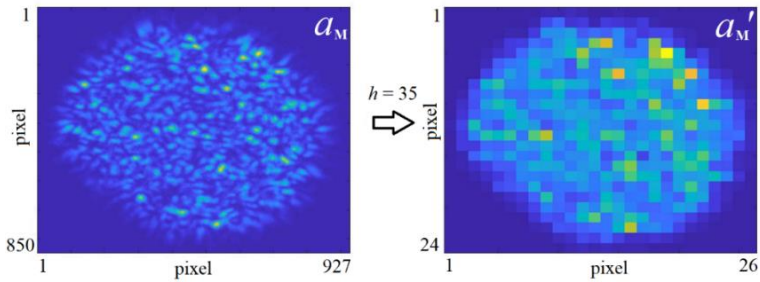

Fig. 4. Cropped original camera image of speckle patterns ($a_M$, left-hand side) compared to the processed mask pattern ($a_M'$, right-hand side) with scaling factor h=35 pixels, which corresponds with the mean speckle size of patter $a_M$.

#### 4.2.1 Mask rescaling effect

By using a set of simulations, we examined how the image reconstruction quality is affected by a varying mask scaling factor. The set of masks employed in the simulations, i.e. camera images of speckles, was acquired in the real measurements. The mean speckle size $h$ for the examined set of masks is $h = 35$ pixels, which was calculated as full width half maximum (FWHM) of the speckle pattern autocorrelation function [20]. The compression ratio $k$ was set for the purpose of simulations to 0.4, and the noise level of the PL intensity was set to the $\sigma = 0.5\%$.

The reconstruction error of the PL maps was calculated as an $l_2$ norm of the vectorized reconstructed image $m$ and the original image $U$. To normalize the error for the image intensity, we calculate the relative error $r$:

$$r = \frac{\|m - U\|_2}{\|U\|_2} \quad (6)$$

Since the scaling factor changes the number of pixels of a mask, the number of reconstructed image pixels $N$ changed accordingly. The straightforward evaluation of the reconstruction quality by using residues was not meaningful because a smaller number of pixels leads to a lower level of residues despite worse image quality since we solve a highly underdetermined system.

The simulations based on two different PL maps in Fig. 5A-B show that the relative error $r$ does not have a strong systematical dependence on the scaling factor (see Fig. 5C). It is only possible to observe a slightly decreasing trend of the error towards smaller scaling values. However, maintaining a given compression ratio for higher resolution implies increasing the number of scanned masks, thus increasing the total acquisition time. Therefore, as a reasonable compromise, we used the scaling factor $h = 35$ that corresponded to the average speckle size of the used patterns. The effect of this scaling is illustrated in Fig. 4, which was processed based on this value.

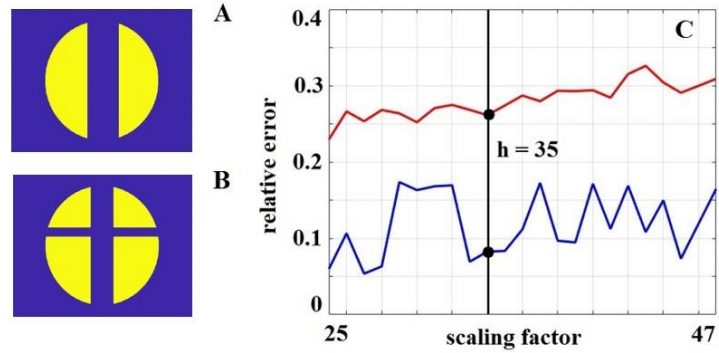

Fig. 5. Effect of scaling factor **h** on relative error in image reconstruction of two different PL maps depicted in panel A (red line in C) and panel B (blue line in C). The black line in panel C denotes the scaling factor according to the mean speckle size.

*4.3. Proof-of-principle measurements*

The first analyzed sample was an orange absorbing cut-off filter OG565, which was divided with an opaque line into two regions with the same PL decay dynamics $I_D$. Such a situation corresponds, for instance, to a mapping of a single PL marker in a sample. The illuminated spot was sized ~ 18 mm², the number of masks $M = 400$. Mask resolution was rescaled according to speckle size ($h = 35$ pixels), leading to the image resolution of 28x36 ($N = 1008$).

We tested image reconstruction for three different compression ratios $k$, where the number of pixels $N$ remained the same and the number of used masks $M$ was decreased accordingly. Namely, we employed the compression ratio $k = 0.4$ (see Fig. 6A), $k = 0.2$ (see Fig. 6B), and $k = 0.05$ (see Fig. 6C). The corresponding data acquisition times were 47 min, 24 min, 6 min, respectively. The results are summarized in Fig. 6 divided into areas A, B and C accordingly.

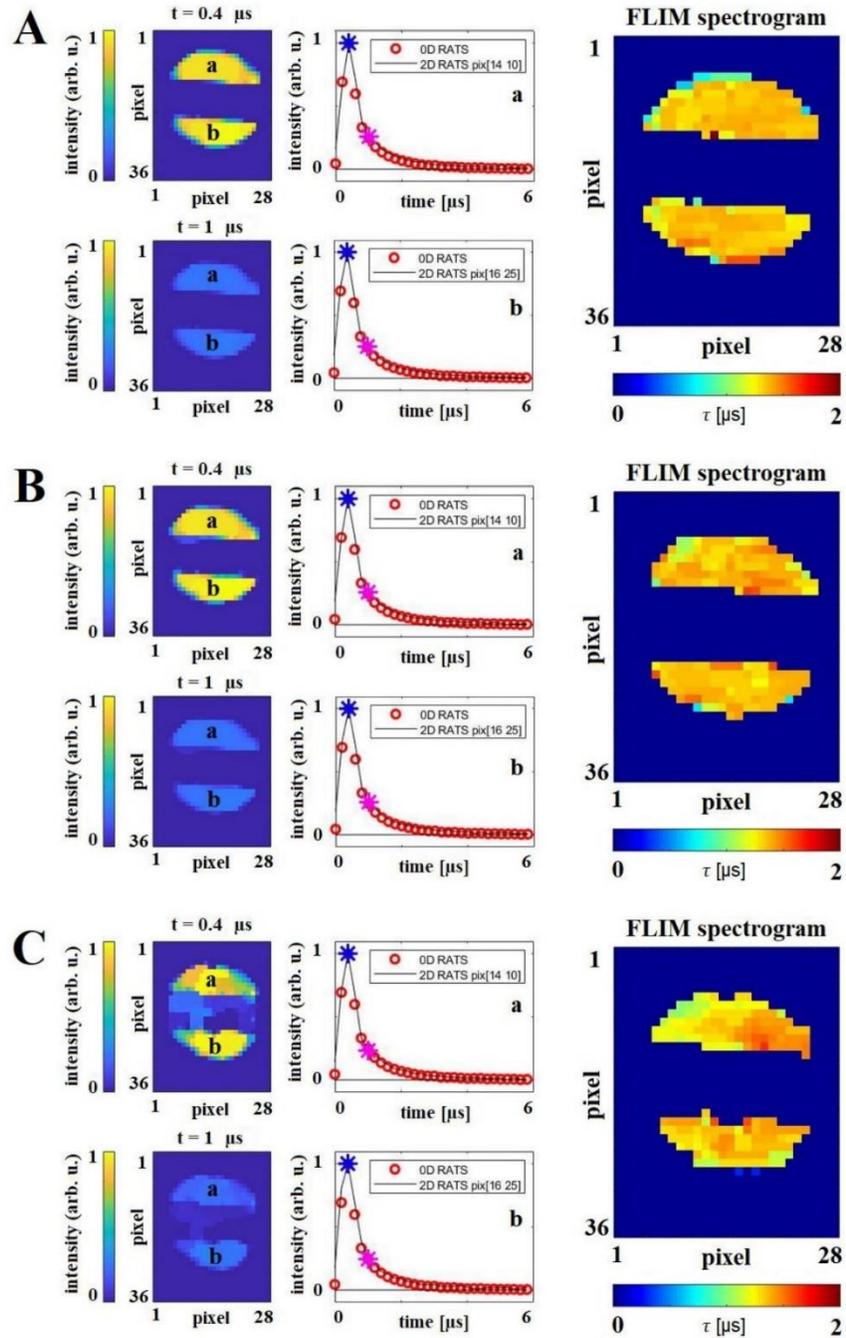

Fig. 6: Measurement of a masked OG565 filter: (A) compression ratio 0.4, (B) compression ratio 0.2, (C) compression ratio 0.05. The left panels - reconstruction of the PL map for two different times. The middle part - graphs "a" and "b" shows the $I_D$ in a randomly selected pixel, which corresponds to the reconstructed area "a", respectively "b"; blue and violet stars denote a time of the PL maps on the left. The right part: map of the PL lifetimes for the points, where the PL amplitude exceeded 10% of the maximum PL intensity.

The left part shows the reconstruction of the PL map for two different times. The middle part includes graphs "a" and "b", which show the $I_D$ of a randomly selected pixel, which corresponds to the reconstructed area "a", respectively. "b". The graphs include two time points (blue and violet), which correspond with time points of the PL maps from the left part of the figure. The reconstructed $I_D$ data (lines) were compared to the 0D RATS method (red circles). In all cases, the PL decays obtained by the 2D RATS method are in perfect agreement with data from the 0D RATS method. Although the reconstructed PL maps for compressed ratio $k = 0.05$ are noisy compared to the higher compression ratios, the PL decays from 0D and 2D RATS methods are still in perfect agreement.

The FLIM spectrogram is shown on the right side of the image. Individual $\tau$ values were determined with the fitting algorithm. The reconstructed curves of $I_D$ were in each pixel fitted by the sum of two exponentials. Lifetime $\tau$ was then determined as the time when the intensity of $I_D$ drops to 10% of the maximum. The impulse response function of measurement was 0.47 μs. The mean lifetimes for the sample measured with compressed ratios 0.4, 0.2, 0.05 are 1.31 μs, 1.29 μs, 1.29 μs. Mentioned average lifetimes vary with standard deviations of 0.09 μs, 0.10 μs, 0.13 μs. The statistical data does not include points that did not show a luminescence intensity lower than 10% of the sample maximum, as well as data from sample edge that were suffering from scattering signal and high noise level. We observed that the lifetime precision, i.e. acquired standard deviations, are only marginally affected by the used compression ratio. This ratio rather affects the quality of the lifetime maps.

In the second measurement, we acquired FLIM data of an artificially prepared sample with two different dynamics of PL decays $I_D$. The first area was an OG565 color filter, and the second area consisted of a Si wafer with a nanoporous surface. Both OG565 filter and nanoporous Si were previously tested with a standard method and the method 0D RATS, and the results from both 0D RATS methods (Fig. 7, red crosses) were compared in randomly chosen pixels of 2D PL map (see Fig. 7 solid lines). The results are summarized in Fig. 7 which has got the same logic as Fig. 6, but the compressed ratios are different.

The measured area was around 18 mm$^2$; in total 600 masks were scanned. Mask resolution was rescaled again according to the mean speckle size (35 pixels), leading to the resolution of the reconstructed PL map of 26x24 ($N = 624$). Due to the lower $I_{PL}$ amplitude of nanoporous Si, the reconstructed data suffer from a lower signal-to-noise ratio. For this reason, we present reconstructed data for compressed ratios k=0.9 (Fig 7A), k=0.7 (Fig 7B), k=0.5 (Fig 7C). The corresponding data acquisition times were 63 min, 49 min, 35 min, respectively. The PL maps in Fig. 7 (left-hand side) were normalized so, that each data point has the same amplitude. Therefore, we observe a flat PL image at the early times (0.6 μs). While for the later time (1.6 μs), the prominent PL intensity is emitted from the upper part, i.e. nanoporous Si with a long PL decay.

Analogously to the previous measurement, individual $\tau$ values were determined again by fitting the data with double-exponential decay for both areas (OG565 and nanoporous Si). Lifetime $\tau$ corresponds again to the time, where $I_D$ intensity drops to 10% of the amplitude. The impulse response function of measurement was 0.47 μs. For the nanoporous Si wafer, the mean lifetimes for the compressed ratios of 0.9, 0.7, and 0.5 corresponded to 21 μs, 21 μs, 20 μs with a standard deviation of 3 μs, 3 μs, 4 μs, respectively. For the OG565 filter, the mean lifetimes were 1.19 μs, 1.24 μs, 1.28 μs varying with a standard deviation of 0.09 μs, 0.16 μs, 0.27 μs. The statistics included again only points that did not have a luminescence intensity greater than 10% of the sample maximum, as well as the edge points of the sample with prevailing scattering signal. In the combined sample, we attained for all measurements a slightly lower lifetime of the OG565 filter area compared to the first sample. This arises due to the highly scattering Si wafer, which leads to a stronger leakage of excitation signal compared to the first measurement. Subsequently, more scattered excitation light gets to the detector and influence the results, because it forms a response-function-limited peak, which is not possible to completely separate from the PL decay.

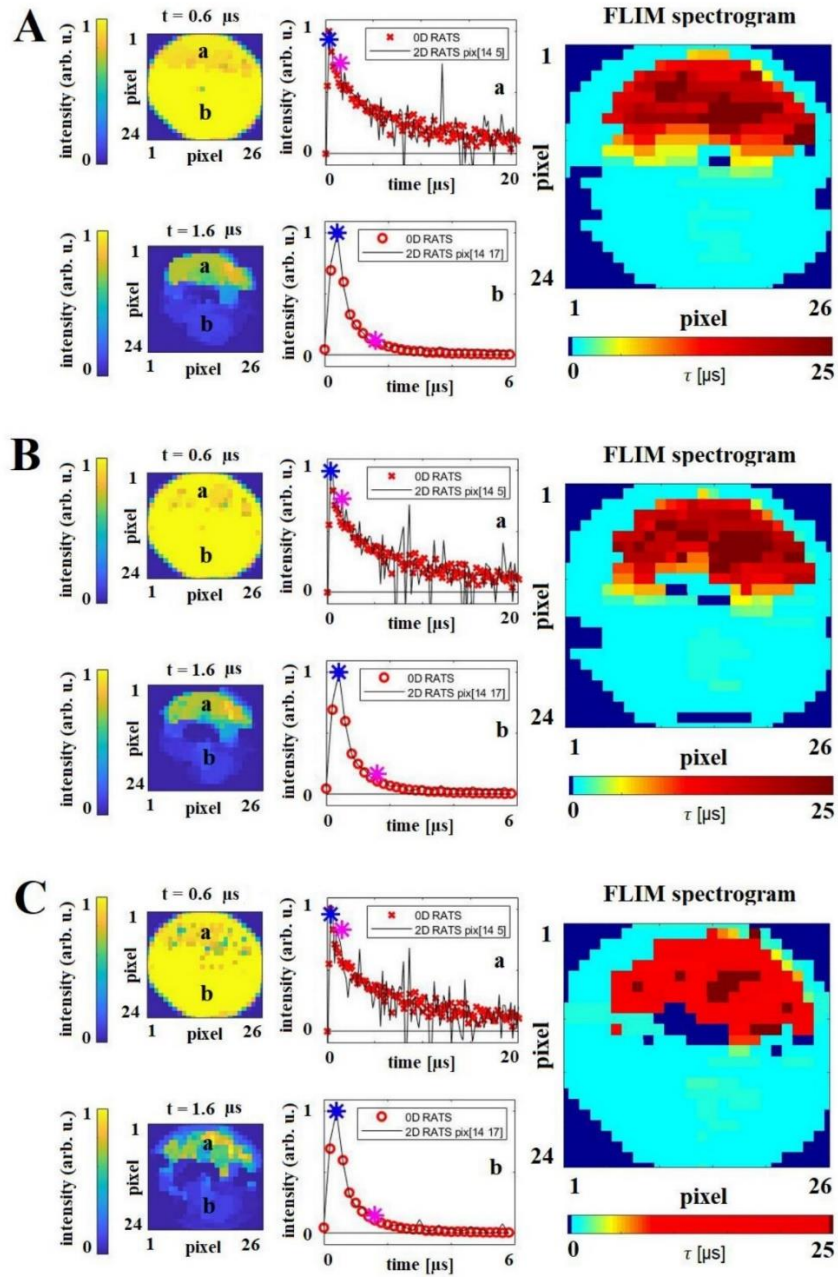

Fig. 7: Measurement of nanoporous Si wafer. (A) compression ratio 0.9, (B) compression ratio 0.7, (C) compression ratio 0.5. The left panels - reconstruction of the PL map for two different times. The middle part - graphs "a" and "b" shows the $I_D$ in a randomly selected pixel, which corresponds to the reconstructed area "a", respectively "b"; blue and violet stars denote a time of the PL maps on the left. The right part: map of the PL lifetimes for the points, where the PL amplitude exceeded 10% of the maximum PL intensity.

## 4.4. Reconstruction error vs. intensity of the decay

The intensity of PL is crucial for the resulting data reconstruction. This can be documented by the fact that the attained PL decay of the nanoporous Si suffers from a significantly higher noise level compared to the OG565 filter data. Analogously, we observed that the noise level of the PL decay increases with delay after excitation, as the PL intensity is decaying and decreasing. This effect was studied by using a relative error of reconstruction. The relative error $\sigma$ was determined as the average absolute deviation of the back reconstructed intensity signal $d_0 = Am$ (see Eq. (5)) and the original signal $d$ at a given time point relative to the average value $\bar{d}$:

$$\sigma = \frac{\overline{|d_0 - d|}}{\bar{d}} \qquad (7)$$

In the Fig.8A (top panel) are shown representative reconstructed decays for the first measurement (two areas with the same PL decay (OG565)). The PL decay of the upper part of the measured sample is indicated by a red line and the lower part with a blue line. In the bottom panel of Fig.8A is evaluated relative error of reconstruction. The same logic is also given in Fig. 8B, which shows representative decays for the combined sample (nanoporous Si + OG565 filter). The red line shows the PL decay of the upper part of the sample (nanoporous Si) and the blue line shows the PL decay of the lower part of the sample (filter OG565). The bottom panel of Fig. 8B then shows the relative reconstruction error. The relative error comparison according to Eq. (7) depicted in Fig. 8 (bottom panels) was done for the same compression ratio $k = 0.6$ and the same image resolution 26x24. This ensures that the length of the vector d is always the same. For the cases, where the length of the vector d changes, it is more appropriate to observe the reconstruction error with Eq. (6), because a lower number of elements of d can cause the reconstruction algorithm TVAL3 to reach a better agreement between d and $d_0$, while the reconstruction of the PL map can feature a lower quality.

For both cases in Fig. 8, we observe that the relative error level steadily increases with the decreasing PL intensity (see the bottom panels). Thus, we can conclude that for a higher intensity of detected PL it would be possible to reconstruct decays with a lower relative error even for small compression ratios. However, the relative error still increases with the decreasing intensity of a PL decay.

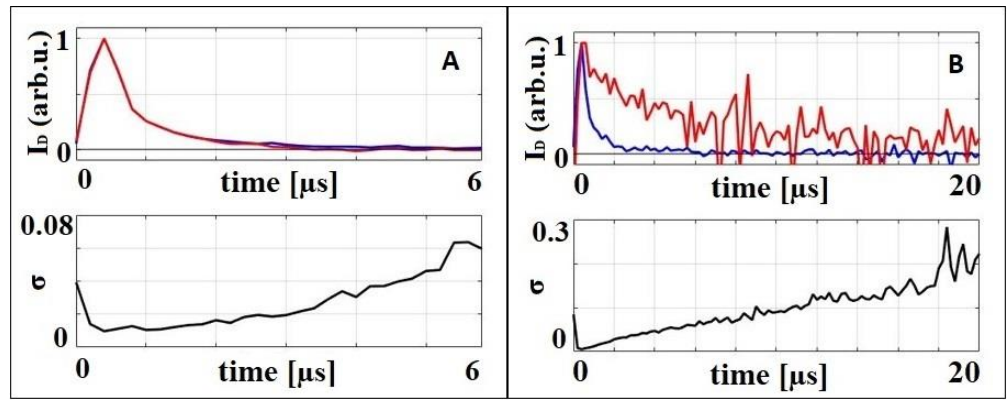

Fig. 8. Reconstruction error evaluation regarding the decreasing intensity of $I_D$. Panel A describes the sample with two the same areas (filter OG565) and panel B describes the sample with two different areas (nanoporous Si + filter OG565) – see text for details.

## 5. Conclusion

We present a new approach to FLIM, which is based on a combination of random temporal speckles (RATS) method together with the concept of a single-pixel camera. Spatio-temporal random speckle patterns make it possible to track both PL images and dynamics on the microsecond timescale. The speckle generation is based on rotating and movable diffusers, which makes the method a low-cost FLIM setup with unmatched simplicity. The strength of the new concept lies in imaging PL decays in the microsecond timescale because the acquisition time is highly reduced compared to the standard FLIM setups employing TCSPC.

The acquisition time depends on several factors. The most prominent one is the PL map resolution and the connected compression ratio. We showed that it is reasonable to use scaling according to the mean speckle size. A lower resolution causes a loss in the image quality, while a higher resolution cannot provide more information, as the resolution is limited by the mean speckle size. The FLIM data reconstruction error also highly depends on the intensity of PL. Hence, we can reach a fast acquisition by using a higher PL intensity, while decreasing a compression ratio. Under ideal conditions, i.e. highly emitting samples, we were able to reach an acquisition time of 6 minutes. For a standard sample, it was required to increase the compression ratio and the resulting acquisition time reaches 35 minutes.

Since this article serves as a demonstration of the new method, further optimization of the optical setup can provide us with more efficient excitation use and PL collection. In summary, the method is a low-cost and straightforward alternative to commonly-used methods, providing the possibility of fast measurement of fast mapping of PL decays on the microsecond timescales.

**Funding**

The Czech Academy of Sciences, ERC-CZ/AV-B (RUSH, Reg. No. ERC100431901); Ministry of Education, Youth and Sports ("Partnership for Excellence in Superprecise Optics," Reg. No. CZ.02.1.01/0.0/0.0/16_026/0008390), The Student Grant Competition of the Technical University of Liberec (under the project No. SGS-2020-3057); Grant Agency of the Czech Republic (Project 17-26284Y).

**Acknowledgment**

We gratefully acknowledge Lukáš Ondič (Institute of Physics, The Czech Academy of Sciences) for providing us with samples of nanoporous Si.

**Disclosures**

The authors declare no conflicts of interest.